# Realization of Practical Eightfold Fermions and Fourfold van Hove Singularity in TaCo$_2$Te$_2$

Hongtao Rong[1*], Zhenqiao Huang[1,7*], Xin Zhang[2,8*], Shiv Kumar[3*], Fayuang Zhang[1], Chengcheng Zhang[1], Yuan Wang[1], Zhanyang Hao[1], Yongqing Cai[1], Le Wang[1], Cai Liu[1], Xiao-Ming Ma[1], Shu Guo[1], Bing Shen[4], Yi Liu[5], Shengtao Cui[5], Kenya Shimada[3], Quansheng Wu[6,10], Junhao Lin[1], Yugui Yao[2,8], Zhiwei Wang[2,8,9#], Hu Xu[1#], and Chaoyu Chen[1#]

[1] Shenzhen Institute for Quantum Science and Engineering (SIQSE) and Department of Physics, Southern University of Science and Technology (SUSTech), Shenzhen 518055, China.

[2] Centre for Quantum Physics, Key Laboratory of Advanced Optoelectronic Quantum Architecture and Measurement, School of Physics, Beijing Institute of Technology, Beijing 100081, China.

[3] Hiroshima Synchrotron Radiation Centre, Hiroshima University, Higashi-Hiroshima, Hiroshima 739-0046, Japan.

[4] School of Physics, Sun Yat-Sen University, Guangzhou 510275, China.

[5] National Synchrotron Radiation Laboratory, University of Science and Technology of China, Hefei, Anhui 230029, China.

[6] Beijing National Laboratory for Condensed Matter Physics, and Institute of Physics, Chinese Academy of Sciences, Beijing 100190, China.

[7] Department of Physics, The Hong Kong University of Science and Technology, Clear Water Bay, Hong Kong, China.

[8] Beijing Key Lab of Nanophotonics and Ultrafifine Optoelectronic Systems, Beijing Institute of Technology, Beijing 100081, China.

[9] Material Science Center, Yangtze Delta Region Academy of Beijing Institute of Technology, Jiaxing 314011, China.

[10] University of Chinese Academy of Sciences, Beijing 100049, China.

* These authors contributed equally to this work.

#Correspondence should be addressed to Z.W. (zhiweiwang@bit.edu.cn), H.X. (xuh@sustech.edu.cn) and C.C. (chency@sustech.edu.cn)

**Space groups describing the symmetry of lattice structure allow the emergence of fermionic quasiparticles with various degeneracy in the band structure. Theoretical efforts have predicted many materials hosting fermions with the highest degeneracy, *i.e.*, eightfold fermions, yet lacking experimental realization. Here, we explore the band degeneracies in TaCo$_2$Te$_2$ crystals. Through systematic experimental and theoretical analyses, we establish TaCo$_2$Te$_2$ as a nonsymmorphic crystal with negligible spin-orbit coupling (SOC) and long-range magnetic order. These critical properties guarantee the first realization of practical eightfold fermions and fourfold van Hove singularity, as directly observed by photoemission spectroscopy. TaCo$_2$Te$_2$ serves as a topological quantum critical platform, which can be tuned into various magnetic, topologically trivial, and nontrivial phases by adding strain, magnetic field, or SOC. The latter is demonstrated by our first-principles calculations, which show that enhancing SOC in TaCo$_2$Te$_2$ will promote the experimental observation of bulk hourglass fermions. Our results establish TaCo$_2$Te$_2$ as a unique platform to explore states of matter intertwining magnetism, correlation, symmetry, and band topology.**

It has been an eternal theme in condensed matter physics to explore emerging low-energy elementary excitations from many-body interactions and/or symmetry constraints. In analog to high-energy physics where the Pioncaré symmetry allows relativistic massless Weyl and Dirac fermions, in condensed matter physics Weyl and Dirac type low-energy quasiparticle excitations have been realized based on materials like graphene [1,2], topological insulator [3-6] and topological semimetals [7-14]. At zero momentum, Weyl and Dirac fermions are twofold and fourfold degenerate, respectively. Moreover, as subgroups of the Pioncaré symmetry, the 230 space groups in condensed matter physics impose fewer constraints on the allowed types of fermions. New fermionic quasiparticles beyond high-energy physics, including threefold, sixfold, and eightfold fermions, can emerge as band degeneracy [15-26]. Experimental exploration of such band degeneracy represents a major and yet-to-be accomplished task.

Table I: Experimental observation of different types of fermions in material families.

| order of dispersion $n$ | band degeneracy $g$ | | | | |
|---|---|---|---|---|---|
| | 2 | 3 | 4 | 6 | 8 |
| 1 | TaAs [7-11], MoTe$_2$ [27-29], LaAlGe [30], TaIrTe4 [31], YbMnBi$_2$ [32], Co$_2$MnGa [33], Co$_3$Sn$_2$S$_2$ [34] | MoP [35,36] | Na$_3$Bi [13,14] Cd$_3$As$_2$ [12] PtSe$_2$ [37-42] | CoSi [43-47] PdBiSe [48] | TaCo$_2$Te$_2$ This work |
| 2 | ? (awaiting experimental proof) | ? | α-Sn [49] | PdSb$_2$ [50-52] | Forbidden |
| 3 | ? | Forbidden | ? | Forbidden | Forbidden |

In Table I, the spectroscopic observation of fermionic excitations as band degeneracy in crystalline materials is summarized according to the band degeneracy $g$ and the order of dispersion $n$. According to theoretical analysis, there exist cubic ($n = 3$) Weyl and Dirac fermions [22,24] yet lacking experimental observation. Quadratic ($n = 2$) Dirac and sixfold fermions have been observed directly by angle-resolved photoemission spectraoscopy (ARPES) in **α**-Sn [49] and PdSb$_2$ [50-52], respectively. Besides that, all the experimentally realized fermionic quasiparticles are of linear dispersion ($n = 1$) with band degeneracy $g$ = 2, 3, 4, 6. However, for quasiparticles with the highest band degeneracy, *i.e.*, eightfold fermions, despite its theoretical prediction based on space groups 130, 135, 218, 220, 222, 223, 230 [16,20,22,25,26], the spectroscopic observation is still missing. This is particularly due to the difficulty in growing or handling the candidate materials [53-55]. Our approach to overcome this is to seek materials with negligible spin-orbit coupling (SOC) effect, which looses symmetry constraints and expands the space group selection.

In this work, we report the practical realization of eightfold fermion in a nonsymmorphic (space group 62) crystal TaCo$_2$Te$_2$. Magnetic property measurements reveal no signature of long-range magnetic order despite the existence of Co atoms. ARPES measurement revealed a double Dirac cone (DDC) feature at the Brillouin zone (BZ) $\bar{X}$ point. Density functional theory (DFT) analysis prove this feature as an eightfold degenerate node at the $S$ point of bulk BZ, protected by the combination of crystalline and time-reversal symmetry. Inclusion of SOC only opens a negligible gap, rendering a practical realization of eightfold fermions. Such symmetry combination also protects a fourfold (robust against SOC) van Hove singularity (VHS) at $Z$ close to the Fermi level ($E_F$), which represents a new type of VHS beyond the recent classifications of high-order VHSs [56,57]. Both the eightfold fermion and fourfold VHS have been directly observed by ARPES. These magnetic and electronic properties establish TaCo$_2$Te$_2$ as a topological quantum critical platform that can be tuned into various magnetic, topologically trivial, or nontrivial phases [16,20], say, by inducing magnetism, SOC, symmetry breaking via chemical, strain, or band engineering. As an example, we use DFT calculation to prove that enhancement of SOC would promote practical ARPES observation of bulk hourglass fermions for the first time. These results suggest TaCo$_2$Te$_2$ as a unique platform to explore emergent physics intertwining magnetism, correlation, symmetry, and band topology.

TaCo$_2$Te$_2$ crystallizes in an orthorhombic structure [58]. Using single-crystal X-ray diffraction (SCXRD) we determine its lattice structure as lattice constants $a = 6.6116\,\text{Å}, b = 6.5735\,\text{Å}, c = 17.783\,\text{Å}$ in space group 62 (*Pmcn*). The corresponding BZ is shown in Fig. 1(b). The crystallinity of as-grown single crystals was examined by XRD. As shown in Fig. 1(c), all the diffraction peaks can be well indexed by the (00l) reflections with $c = 17.783\,\text{Å}$. The magnetic property of TaCo$_2$Te$_2$ is shown in Fig. 1(d). First, no order behaviour can be observed from $\chi - T$ measurement, indicating the absence of magnetic order in the current temperature region. Between 100 K and 300 K, the slope of $\chi - T$ curves can be positive or negative depending on samples, neither of which can be fitted by Curie-Weiss law, indicating extrinsic origin. Secondly, the value of $\chi$ in TaCo$_2$Te$_2$ is 3 orders of magnitude lower compared to typical cobalt compound Co$_3$Sn$_2$S$_2$ [59,60], suggesting negligible Co moment. Thirdly, the field-dependent magnetization ($M - H$) curves also show no signature of saturation or hysteresis. All the above magnetic properties enable us to omit the magnetism of TaCo$_2$Te$_2$ in the following electronic structure study.

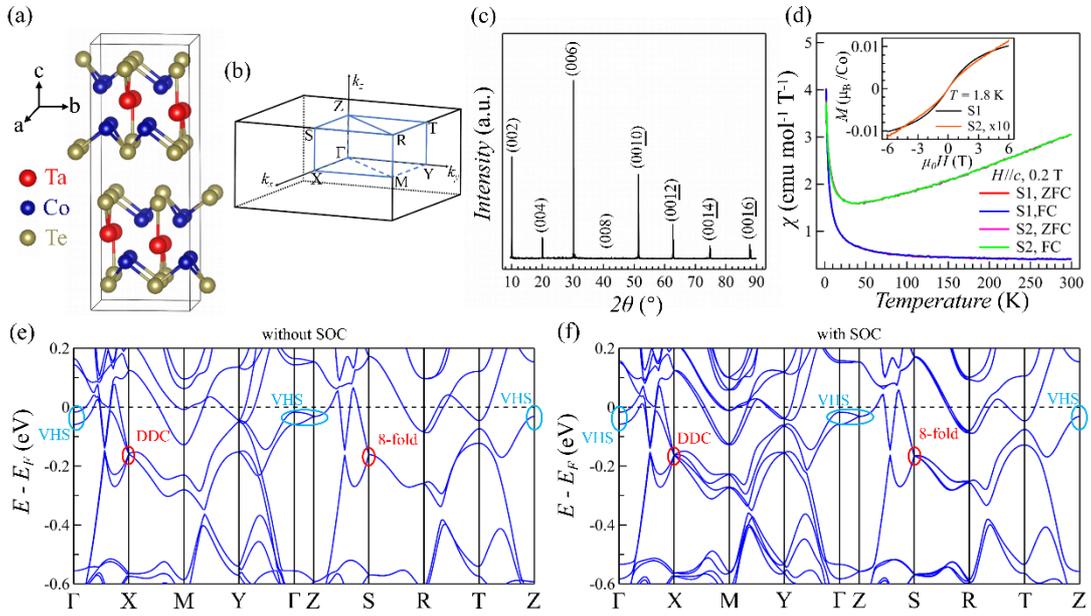

**FIG. 1. Eightfold fermions and fourfold VHS in nonsymmorphic TaCo$_2$Te$_2$ with negligible magnetic order.** (a) Crystal structure of TaCo$_2$Te$_2$. (b) Three-dimensional BZ of TaCo$_2$Te$_2$. (c) Single crystal x-ray diffraction (XRD) result taken at 300 K. (d) Zero-field-cooled (ZFC) and field-cooled (FC) magnetic susceptibility ($\chi$) vs temperature ($T$) for magnetic field $H = 0.2\,T$ parallel to the $c$ axis. Inset: Field-dependent magnetization ($M$) at 1.8 K. (e, f) DFT calculated dispersions along the high-symmetry path without considering SOC (e) and with SOC (f).

Figures 1(e) and 1(f) show the band structure of TaCo$_2$Te$_2$ along the high-symmetry path calculated by DFT without and with considering SOC, respectively. Two exotic features in the $k_z = 0$ plane ($\Gamma - X - M - Y - \Gamma$ plane) are noticeable. The first one is a DDC feature ($\sim -0.17\,eV$) at $X$ point (highlighted by red ellipses). Fig. 2(g) shows that this DDC has two fourfold degenerate points separated by a Dirac gap $\sim 0.7\,meV$. The second one is a double VHS feature close to $E_F$ at $\Gamma$ point (highlighted by blue ellipses). Along $\Gamma - X$ direction they are electron pockets centered at $\Gamma$ and crossing $E_F$, while along $\Gamma - Y$ they are hole-like, yielding two saddle point VHSs.

When moving to $k_z = \pi$ plane (Z − S − R − T − Z plane), the energy splitting between band pairs decreases and even vanishes, resulting in doubled band degeneracy. The DDC feature at $X$ merges into a single, fourfold Dirac cone with a gapless, eightfold degenerate point at $S$ (Fig. 2(h-i)). This eightfold fermion is symmetry-protected without SOC, determined by the projective irreducible representations (IReps) of the little group at $S$. Since the system is nonmagnetic without considering SOC, we only need to consider single-valued IReps. The nonsymmorphic group $Pmcn$ has higher dimension IReps at the boundary of the first BZ [61]. At $S$, the little group has two two-dimension IReps, which are conjugated to each other. Due to the time reversal symmetry, the two conjugated IReps will become a four-dimension IRep through the direct sum. Moreover, since the SOC here is negligible, the spin space has SU(2) symmetry protecting the spin degeneracy. Therefore, the band at the point S is eightfold degenerate. On the other hand, when considering SOC, the degeneracy will be determined by the double-valued projective IReps of the little group and the SU(2) symmetry will be broken. Although the double-valued IReps of the little group at $S$ also include a four-dimension IRep, the time reversal cannot lead to eightfold degeneracy [16,20]. Therefore, SOC will generally open a gap at $S$. But the following ARPES measurement (unresolvable) and DFT analysis ($\sim 0.4\ meV$) both manifest that this SOC induced gap is negligible.

Besides, the double VHSs at $\Gamma$ merge into one single VHS at $Z$, forming a fourfold VHS, robust against SOC. First, $Z$ is invariant under inversion symmetry $P$ and a glide mirror $\widetilde{M}_x: (x, y, z) \to (-x, y + \frac{1}{2}, z + \frac{1}{2})$. In this case, if a Bloch state $|u\rangle$ is an eigenstate of $\widetilde{M}_x$ with eigenvalue $g_x$, there is another state of the system $P|u\rangle$. Moreover, because of the fractional translation of $\widetilde{M}_x$, the two operators are anticommutative, $\{P, \widetilde{M}_x\} = 0$, which gives $M_x P|u\rangle = -P\widetilde{M}_x|u\rangle = -g_x P|u\rangle$. Thus, $|u\rangle$ and $P|u\rangle$ must be linearly independent for their opposite eigenvalues. Then, since $Z$ is a time-reversal invariant momentum and the time reversal operator $T$ has $T^2 = -1$, the band at this point has a Kramer double degeneracy. Therefore, for the states $|u\rangle$ and $P|u\rangle$, there must be linearly independent states $T|u\rangle$ and $TP|u\rangle$ having the same energy. Because $PT = TP$ here, $TP|u\rangle$ is equal to $PT|u\rangle$. So, the independence of $T|u\rangle$ and $TP|u\rangle$ can be also argued by the anticommutation relation. Furthermore, the $Z$ point is also invariant under the combination of time reversal and inversion symmetries $TP$ with $(TP)^2 = -1$. Thus, $TP|u\rangle$ is another Kramer-like double degenerate state of $|u\rangle$ at $Z$. Therefore, for any eigenstates $|u\rangle$ of $\widetilde{M}_x$ with momentum vector $(0,0,\pi)$ in the crystal, there must be three linearly independent states $P|u\rangle, T|u\rangle$, and $TP|u\rangle$. Namely, the bands at $Z$ are all fourfold degenerate.

Photon energy dependent ARPES measurement shown in Fig. 2(a) presents a periodic pattern of spectral intensity in the $k_x - k_z$ plane taken at $E_B = 0.13\ eV$, helping to define the high symmetry points of $\Gamma, Z, X, S$, using lattice constant $c = 17.78\ \text{Å}$ and inner potential $V_0 = 13.5\ eV$. Fig. 2(e) shows the corresponding band dispersion along $k_z$ direction at $k_x = 0.67\ \text{Å}^{-1}$. Fermi surfaces mapped by ARPES at different $k_z$ planes in Fig. 2(b) and 2(c) present features consistent with the DFT projection (Fig. 2(d)). Of particular interest are two elliptical features next to $\bar{X}$, which come from the DDC features along $\bar{\Gamma} - \bar{X} - \bar{\Gamma}$ direction. In Fig. 2(f-i), we compare the ARPES measured DDC spectra along $\Gamma - X - \Gamma$ and $Z - S - Z$ to the DFT calculated ones. For the DDC feature centered at $X$, the band pair splitting can be clearly resolved in the 2$^{nd}$ derivative plot and splitting bands are marked by yellow arrows shown in Fig. 2(f). It is noted that this splitting in ARPES spectra is more evident for the outer cones than the inner ones, probably resulting from the fact that the inner ones are more dispersed along $k_z$ direction and its projection into the $k_x - k_y$ plane may smear out its own splitting. According to the above symmetry analysis, moving to $k_z = \pi$ plane merges the DDC feature into a single but fourfold degenerate Dirac cone, with an eightfold degenerate Dirac point (Fig. 2(i)). This is indeed observed in the ARPES spectral along $Z - S - Z$ direction as shown in Fig. 2(h). Compared to ARPES spectra along $\Gamma - X - \Gamma$, spectra along $Z - S - Z$ show the absence of band pair splitting and the presence of a degenerate point with enhanced spectral weight. These new features clearly prove the fourfold band degeneracy and the existence of eightfold fermion. The SOC induced gap (~$0.4\ meV$ according to DFT in Fig. 2(i)) is negligible as compared to the breadth of observed bands in ARPES spectra (Fig. 2(h)).

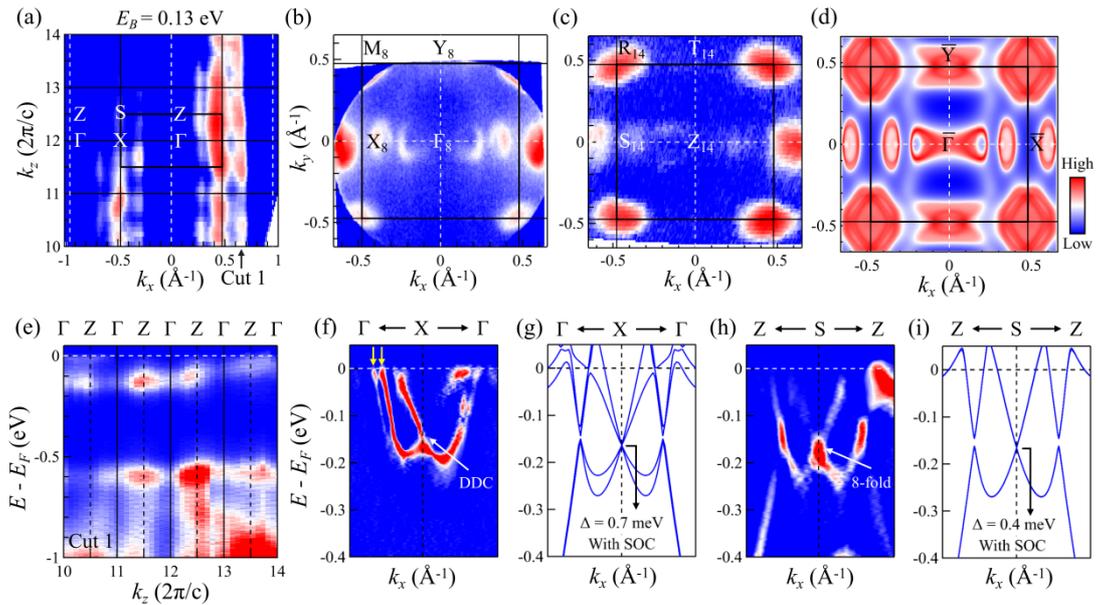

**FIG. 2. Observation of practical eightfold fermions in TaCo₂Te₂.** (a) Integrated photoemission intensity maps in the $k_x - k_z$ plane taken at $E_B = 0.13\ eV$ measured by varying photon energy. (b, c) Fermi surface in the $k_x - k_y$ plane measured with photon energies $24\ eV$ and $90\ eV$, respectively. (d) Fermi surface calculated by DFT with projection to the $k_x - k_y$ plane. (e) The band dispersion along the $k_z$ direction at $k_x = 0.67\ Å^{-1}$ as shown in (a) Cut1. (f-i) ARPES measured spectra (2$^{nd}$ derivative plot) and corresponding DFT calculated dispersions along high-symmetry paths with SOC.

 

We then focus on the fourfold VHS at the $Z$ point as shown in Fig. 3(a). According to DFT, this VHS manifests itself as an electron pocket along $Z - S$ and a hole pocket along $Z - T$, thus as a saddle point with divergent density of states. To measure the corresponding spectra, we first align the $Z - S$ direction parallel to the measurement plane and then make a Fermi surface mapping. As shown by the mapping in Fig. 3(b), the spectral weight is heavily suppressed at $Z$ in the first BZ due to the measurement geometry and orbital symmetry of the contributing electrons. So we choose the extended BZ to extract the cut along $S - Z - S$. As shown in Fig. 3(c), ARPES spectra indeed present an electron-like pocket centered at $Z$ with its minimum binding energy $\sim 32\ meV$. Then the sample is rotated by $90°$ so that $Z - T$ is parallel to the measurement plane. The ARPES spectra along $T - Z - T$ are extracted from the corresponding Fermi surface mapping (Fig. 3(d)). As shown in Fig. 3(e), a hole-like pocket centered at $Z$ can be resolved. Comapring Fig. 3(c) and 3(e). it is found that the minimum of the electron pocket along $S - Z - S$ coincides with the maximum of the hole pocket along $T - Z - T$, being evidence of a saddle point. As our previous symmetry analysis has proved $Z$ as a fourfold degenerate point, this saddle point is thus a fourfold degenerate VHS, which goes beyond the recent classifications of high order VHSs [56,57]. In the energy-resolved density of states (DOS) plot, a significant DOS enhancement at the saddle point energy can be seen (for details, see Fig. S1 [62]).

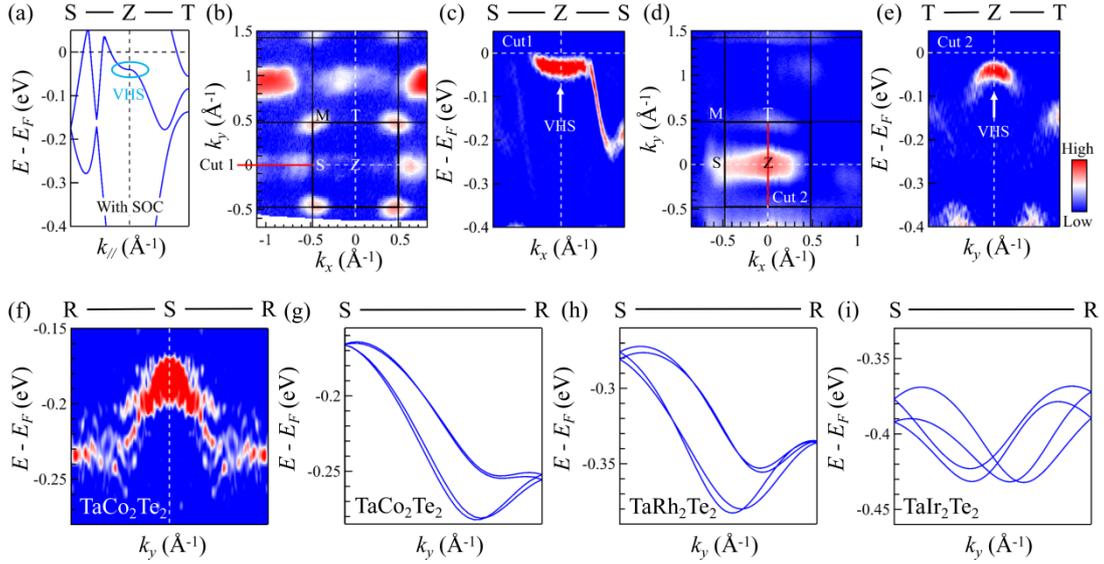

**FIG. 3. Observation of fourfold VHSs in $TaCo_2Te_2$ and predicted realization of hourglass fermions in $TaIr_2Te_2$.** (a) DFT calculated fourfold bands along $S - Z - T$. (b, d) Fermi surface in $k_x - k_y$ plane measured with photon energy $90\ eV$, with the measurement plane defined by the incident light and emitted electrons parallel to $Z - S$ (b) and $Z - T$ (d) directions, respectively. (c, e) 2$^{nd}$ derivative ARPES spectra taken from the momentum cut indicated by red lines in (b) and (d), respectively. (f) 2$^{nd}$ derivative ARPES spectra showing the feature of hourglass fermion predicted in $TaCo_2Te_2$. (g, h, i) DFT predicted hourglass fermions.

**Discussion:**

Combining magnetic, structural, electronic structure experiments and calculations, we have proved that $TaCo_2Te_2$ shows topological quantum critical behaviors related to its magnetic and electronic properties. Chemical substitution or applying strain may change the valence state of Co atoms and introduce large magnetic moments, so that long range magnetic order and magnetic quantum critical point are expected. The negligible SOC leads to the practical realization of eightfold fermions, which, according to theoretical analysis [16,20], serves as a topological quantum critical point. Symmetry breaking via magnetic field or uniaxial strain may lead to various topologically trivial or nontrivial phases such as Dirac point, Weyl point or nodal lines. The enhanced DOS by eightfold degeneracy and fourfold VHS also amplifies electron correlation, potentially resulting in various ordering instabilities and many-body interactions. Thus, we establish $TaCo_2Te_2$ as a conjoint topological and quantum critical platform with handful stimuli available to tune its physical properties.

We take SOC as an example of such a stimulus to theoretically demonstrate its tunability. As shown in Fig. 3(g), the nonsymmorphic symmetry guarantees band features hosting hourglass fermions[18] along $S-R$ in TaCo$_2$Te$_2$, distinct from surface hourglass fermions as discussed in KHgSb [63], which are the surface states in a type of topological insulators[18]. The bulk hourglass fermion in our work is a nodal loop semimetal constructed by the hourglass crossing points in the glide-invariant plane of the BZ [64,65]. It has been shown that the nodal loop in the bulk can give rise to drumhead surface states with a split band at the sample surface where the nodal loop is projected on [66]. As shown in Fig. 3(g), the predicted splitting between adjacent Dirac crossing is too small to be resolved in the ARPES spectra in Fig. 3(f). Increasing the effect of SOC, e.g., via substituting Co with Rh or Ir, would increase the adjacent Dirac crossing splitting and promote the practical observation of hourglass fermion features by ARPES. This expectation is indeed true as proved by our DFT calculation on the isostructural Ta$T_2$Te$_2$, where $T$ represents Co, Rh, and Ir (Fig. S2). As shown in Fig. 3(g-i), in the considered energy-momentum region always exists an hourglass fermion feature for all the three materials. With increasing SOC, the energy splitting between adjacent Dirac points increases from $\sim 3\ meV$ in TaCo$_2$Te$_2$ to $\sim 15\ meV$ in TaIr$_2$Te$_2$. The latter energy scale is comparable to the breadth of observed bands here, thus making the ARPES observation of hourglass fermions and the related drumhead surface states in TaIr$_2$Te$_2$ very promising. Our observation thus calls for further efforts to explore exotic states of matter and emergent phenomena intertwining magnetism, correlation, symmetry, and band topology based on the TaCo$_2$Te$_2$ family of materials.

**ACKNOWLEDGEMENTS**

This work is supported by the National Natural Science Foundation of China (NSFC) (Grants No. 12074163), Guangdong Basic and Applied Basic Research Foundation (Grants No. 2022B1515020046 and No. 2021B1515130007), the Guangdong Innovative and Entrepreneurial Research Team Program (Grants No. 2019ZT08C044), Shenzhen Science and Technology Program (Grant No. KQTD20190929173815000 and No. RCYX20200714114523069). C.C. acknowledges the assistance of SUSTech Core Research Facilities. The work at Beijing Institute of Technology was supported by the Natural Science Foundation of China (Grant No. 92065109), the National Key R&D Program of China (Grant No. 2020YFA0308800), the Beijing Natural Science Foundation (Grant No. Z190006, Z210006). Z.W. thanks the Analysis & Testing Center at BIT for assistance in facility support.

**Materials and Methods**

**Sample growth and Characterizations**

High quality TaCo$_2$Te$_2$ single crystals were synthesized by using a chemical vapor transport method with I$_2$ as a transport agent. The mixture of Ta powder (purity 99.9%), Co pieces (99.99%), and Te shot (99.99%) were prepared with a molar ratio of Ta : Co : Te = 1 : 1 : 2, and sealed in a quartz tube under high vacuum. Then it was placed in a horizon two-zone tube furnace and maintained under a temperature gradient of 950 - 850 °C. After two weeks, shiny single crystals with a typical size of about $1 \times 2 \times 0.1$ mm$^3$ were obtained. The structure of the crystals was characterized by x-ray diffraction with Cu $K\alpha$ radiation at room temperature using a Rigaku Miniex diffractometer. The diffraction pattern can be well indexed by the (00*l*) reflections. Magnetic measurements were performed using the QD Magnetic Property Measurement System (MPMS) with the Vibrating Sample Magnetometer (VSM) mode. Temperature dependent magnetic susceptibility and the field-dependent magnetization results were collected with an external magnetic field of 0.2 T, which is parallel to the *c* axis.

**ARPES measurement**

ARPES measurements were performed at the beamline 13U of the National Synchrotron Radiation Laboratory (NSRL) and the beamline 1 of the Hiroshima Synchrotron Radiation Center (HSRC), Hiroshima, Japan. The energy resolution was set at 25 meV for Fermi surface mapping and 12.5meV for band structure measurements, respectively. The angular resolution was set at 0.1°. Samples were cleaved *in situ* along the (001) crystal plane under ultra-high vacuum with pressure better than $5 \times 10^{-11}$ mbar and temperatures around 20 K.

**First-principles calculations**

The first-principles calculations are performed using the Vienna ab initio simulation package [67,68] within the projector augmented wave method [69]. The exchange correlation functional was treated within the generalized gradient approximation in the Perdew-Burke-Ernzerhof formalism [70]. The plane-wave basis with an energy cutoff of 500 eV. The BZ was sampled by using the Γ-centered 5×5×2 Monkhorst-Pack grid [71] in the self-consistent calculation. The spin polarized calculations are calculated to give a nonmagnetic ground state. In addition, a tight-binding (TB) Hamiltonian based on the maximally localized Wannier functions (MLWF) [72] is constructed by using the Wannier90 package [73]. Through the TB Hamiltonian, the Fermi surface is further plotted by using the WannierTools package [74].

materials. *Computer Physics Communications* **224**, 405-416, doi:https://doi.org/10.1016/j.cpc.2017.09.033 (2018).